\documentclass[12pt]{iopart}
\usepackage{graphicx}
\usepackage{bm}
\usepackage{color}

\def\l{\langle}
\def\r{\rangle}

\begin{document}

\title[BKT transitions on the various two-dimensional lattices]{BKT transitions of the XY and six-state clock models on the various two-dimensional lattices}

\author{Yutaka Okabe and Hiromi Otsuka}

\address{Department of Physics, Tokyo Metropolitan University, Hachioji, Tokyo 192-0397, Japan}
\ead{okabe@phys.se.tmu.ac.jp}
\vspace{10pt}
\begin{indented}
\item[]\today
\end{indented}

\begin{abstract}
In a two-dimensional (2D) spin system, the XY model, characterized by planar rotational symmetry, exhibits a unique phenomenon known as the Berezinskii-Kosterlitz-Thouless (BKT) transition. In contrast, the clock model, which introduces discrete rotational symmetry, exhibits the BKT transition at two different temperatures due to this discreteness.
In this study, we numerically investigate the BKT transition for XY and six-state clock models over various two-dimensional lattices. We employ two primary methods: the Monte Carlo method, which analyzes the size dependence of the ratio of the correlation functions for two different distances, and a machine-learning approach to classify the different phases --- namely, the low-temperature ordered phase, the intermediate BKT phase, and the high-temperature disordered phase.
We identify the BKT transition temperatures for the XY and six-state models on honeycomb, kagome, and diced lattices. 
Combined with the previously calculated data for the triangular lattice, we then compare these values with the second-order phase transition temperatures of the 2D Ising model, for which exact solutions are known. 
Our results indicate that the ratio of the BKT transition temperatures for each lattice relative to the Ising model transition temperatures are close, although the values are not universal. 
\end{abstract}

\begin{indented}
  \item[]Keywords: BKT transition, XY model, clock model, Ising model, Monte Carlo simulation, machine-learning study
\end{indented}

\maketitle

\section{Introduction}

Two-dimensional (2D) spin systems with continuous XY symmetry undergo a unique phase transition known as the Berezinskii-Kosterlitz-Thouless (BKT) transition \cite{Berezinskii1,Berezinskii2,kosterlitz,kosterlitz2}. In the BKT phase, which exhibits quasi-long-range order, the correlation function decays according to a power law. The $q$-state clock model, a discrete counterpart of the XY model, experiences two BKT transitions when $q > 4$ due to its discrete nature \cite{jose,elitzur}. 

However, the numerical analysis of the BKT transition is challenging. The correlation length diverges more rapidly than any power law during the BKT transition, and there are logarithmic corrections~\cite{Janke} that complicate matters. These characteristics make it difficult to accurately determine the BKT transition point through finite-size calculations. 
Monte Carlo simulation is a widely used method for the numerical analysis of many particle systems \cite{Landau}. The finite-size scaling (FSS) study \cite{Barber,Cardy} of the Binder ratio \cite{Binder}, which is essentially a moment ratio, serves as a powerful tool for studying second-order phase transitions. However, the Binder ratio is less effective for analyzing the BKT transition \cite{hasenbusch} due to the presence of multiplicative logarithmic corrections \cite{kosterlitz2,Janke}. Other quantities also exhibit FSS behavior with a single variable, similar to the Binder ratio. Some of these quantities are less sensitive to logarithmic corrections than the Binder ratio. 
For instance, the ratio of correlation functions at different distances \cite{Tomita2002} exhibits this property. In addition, the size-dependent second-moment correlation length for a finite system, denoted as $\xi(L)$, has often been applied to spin glass problems \cite{katzgraber}; the FSS of $\xi(L)/L$ retains the same form with a single scaling variable.
Surungan {\it et al.}~\cite{Surungan} made the Monte Carlo study 
of the correlation ratio and the size-dependent second-moment 
correlation length for the $q$-state clock model.  
They observed the collapsing curves 
of different sizes at intermediate temperatures 
and the spray out at lower and higher temperatures, 
which demonstrates two BKT transitions. 

In addition to conventional approaches, new numerical strategies 
are developing. 
Recent advances in machine-learning-based techniques 
have been applied to fundamental research, such as 
statistical physics \cite{Carleo}. 
Carrasquilla and Melko \cite{Carrasquilla} used a technique 
of supervised learning for image classification, 
which is complementary to the conventional approach 
of studying interacting spin systems. 
They classified and identified a high-temperature paramagnetic phase 
and a low-temperature ferromagnetic phase of 
the 2D Ising model by using data sets of spin configurations. 
Shiina {\it et al.} \cite{Shiina} extended and generalized 
this machine-learning approach to studying various spin models 
including the multi-component systems and the systems 
with a vector order parameter. 
The configuration of a long-range spatial correlation 
was treated instead of the spin configuration itself. 
Not only the second-order and the first-order transitions 
but also the BKT transition was studied. 
They detected two BKT transitions for the six-state clock model. 
Tomita {\it et al.}~\cite{Tomita2020} made further progress 
in the machine-learning study of phase classification. 
When the cluster update is possible in the Monte Carlo simulation, 
the Fortuin-Kasteleyn (FK) \cite{KF,FK} 
representation-based improved estimators 
\cite{Wolff90,Evertz93} for the configuration of two-spin correlation
were employed as an alternative to the ordinary spin configuration.
This method of improved estimators was applied not only 
to the classical spin models but also the quantum Monte Carlo 
simulation. 

In the study of lattices, most investigations of the BKT transition have focused on the square lattice. Highly accurate data on the BKT transition temperature are available for the XY model, as reported by Hasenbusch~\cite{Hasenbusch2005}, and for the $q$-state clock model, as provided by Tomita and Okabe~\cite{Tomita2002b}. 
More recently, tensor network methods have been applied to the study 
of the XY model~\cite{Vanderstraeten} and the $q$-state clock model~\cite{Li}.
The list of the numerical estimates of the BKT temperature of the XY model 
has been tabulated in Table II of Ref.~\cite{Vanderstraeten}, 
and those of the $q$-state clock model in Table S1 
of the Supplemental Material of Ref.~\cite{Li}.
An exception is the study by Sorokin~\cite{Sorokin} for the XY model on a triangular lattice. In addition, the ferromagnetic XY model on a triangular lattice, along with the six-state clock model, was numerically analyzed in the context of the antiferromagnetic Ising model with next-nearest neighbor interactions on a triangular lattice, relating these findings to the six-state clock universality~\cite{Otsuka2023}.
Recently, two research groups~\cite{Andrade,Jiang} have studied the BKT transition of the XY model on a honeycomb lattice, motivated by its relevance to the solution of the $n$-vector loop model~\cite{Nienhuis,Wang_Zhang}. 

In this paper, we perform a systematic study of the BKT transition for both the XY and six-state clock models over various two-dimensional lattices. 
Our calculations include the honeycomb lattice, the kagome lattice, and the diced lattice. For the triangular lattice, we refer to the results discussed in the appendices of a previous study~\cite{Otsuka2023}. 
Exact solutions are known for Ising models on several 2D 
lattices~\cite{Onsager,Husimi,Syozi,Kano}. 
We discuss the relationship between the BKT transition temperatures of the XY and the six-state clock models and the second-order phase transition temperature of the Ising model on various lattices.


The rest of the paper is organized as follows: 
In Sec. 2, we explain the models and simulation methods. 
Section 3 is devoted to the results of the Monte Carlo study 
and the machine-learning study. 
The comparison of the BKT temperatures and the second-order 
phase transition temperatures of the 2D Ising model is discussed 
in Sec. 4.
The summary and discussion are given in Sec. 5.  

\section{Models and simulation methods}

\subsection{Model}
\begin{figure}
\begin{center}
(a) \hspace*{6.4cm} (b)
\vspace*{2mm}

\includegraphics[width=6.2cm]{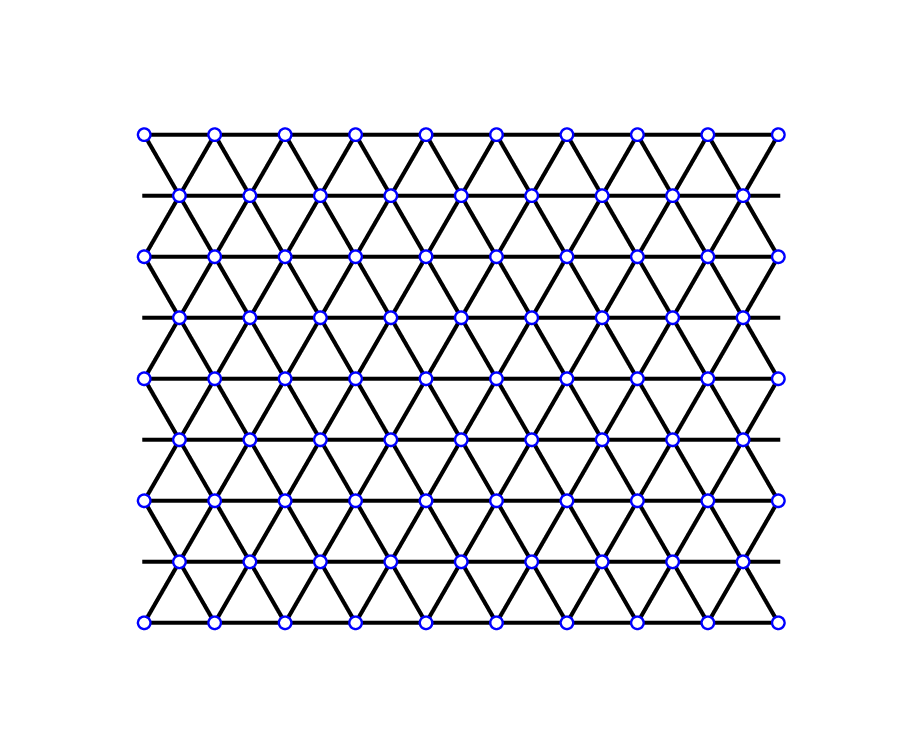}
\hspace*{2mm}
\includegraphics[width=6.2cm]{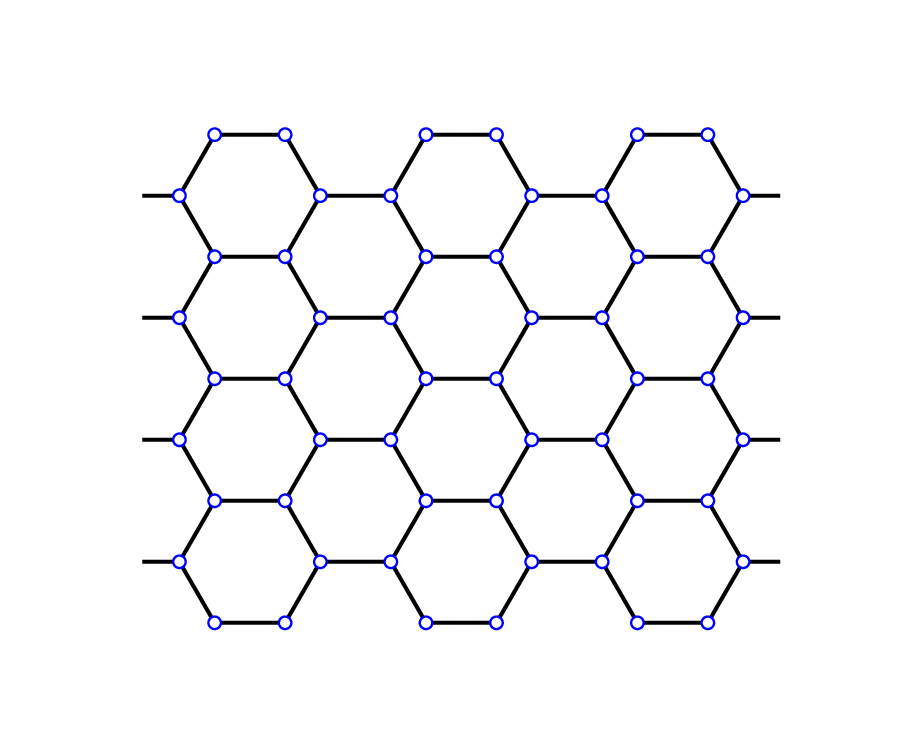}
\vspace*{4mm}

(c) \hspace*{6.4cm} (d)
\vspace*{2mm}

\includegraphics[width=6.2cm]{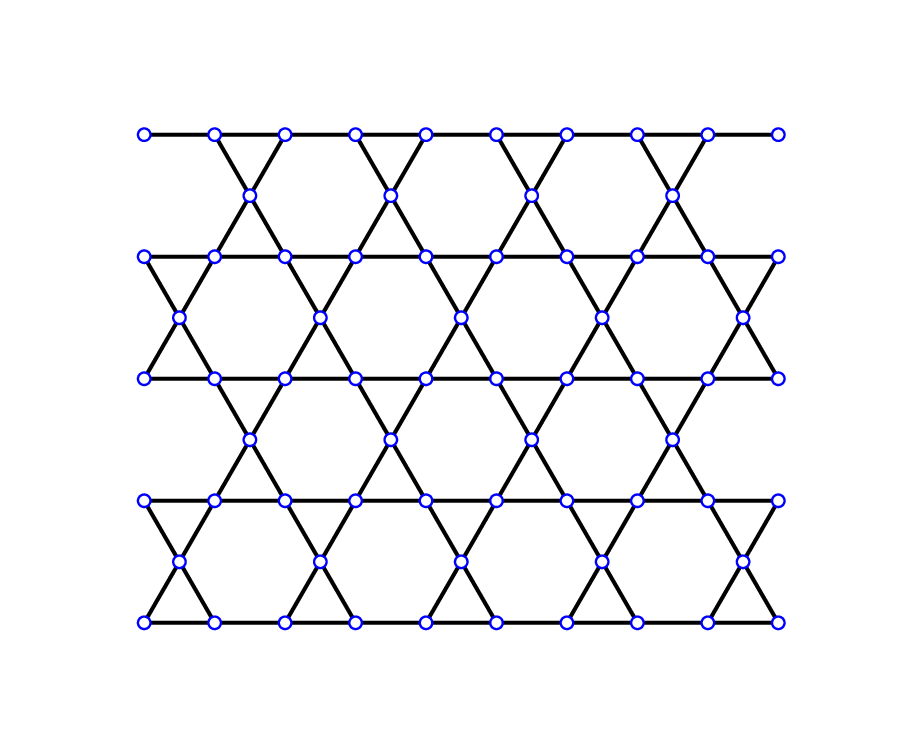}
\hspace*{2mm}
\includegraphics[width=6.2cm]{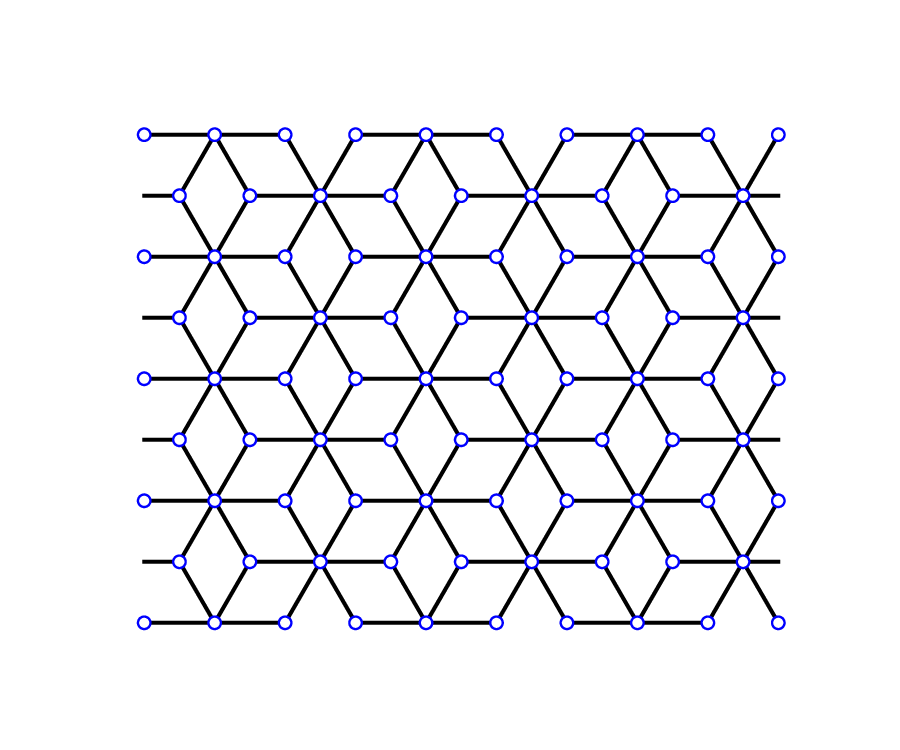}
\caption{
The illustration of (a) triangular, (b) honeycomb, (c) kagome, 
(d) diced lattices.
}
\label{fig:lattices}
\end{center}
\end{figure}

A $q$-state clock model is defined by the following Hamiltonian: 
\begin{equation}\label{Hc}
 H = -J \sum_{\l ij \r} \vec s_i \cdot \vec s_j
   = - J \sum_{\l ij \r} \cos(\theta_i - \theta_j) ,
\end{equation}
where the spins $\vec s_i$ located on lattice sites are planar spins 
restricted to align at $q$ discrete angles ($\theta = 2n\pi/q$ 
with $n = 1,\cdots,q$).  In the limit as $q$ approaches infinity, 
this model becomes an XY model.  
The summation is over the nearest-neighbor sites of 2D lattices. 
We examine the triangular, honeycomb, kagome, and diced lattices, which are illustrated in Fig.~\ref{fig:lattices}. The system sizes considered are \( \frac{3}{2} L \times L \), with periodic boundary conditions imposed for the numerical simulations.

\subsection{Monte Carlo Study}

We utilize two approaches to study the BKT transitions.  
First, we conduct Monte Carlo simulations. 
In this process, we implement the multi-cluster 
spin-flip Swendsen-Wang algorithm~\cite{sw87} and incorporate 
the embedded scheme of Wolff \cite{wolff} for constructing 
clusters of clock spins to overcome the problem of the long 
autocorrelation time near the critical temperature. 
Additionally, we combine the replica exchange method of 
parallel tempering~\cite{Hukushima}. 

We calculate the ratio of the correlation functions of different distances, 
$R(T)=\l g(L/2) \r/\l g(L/4) \r$; for the two distances, 
we chose $L/2$ and $L/4$. 
The correlation function with the distance $r$ is given by
\begin{equation}
    g_i(r) = \vec s_i \cdot \vec s_{i+r} = \cos(\theta_i - \theta_{i+r}).
\end{equation}
The correlation ratio has a single scaling variable 
for FSS
\begin{equation}
  R(T) = \frac{\l g(L/2) \r}{\l g(L/4) \r} = \tilde f(L/\xi),
\end{equation}
as in the Binder ratio~\cite{Binder}. 
Here, $\xi$ stands for the correlation length. 
At the critical region, where the correlation length $\xi$ is infinite, 
the correlation ratio does not depend on the system size $L$. 
Thus, we expect the data collapsing of different sizes in the BKT phase. 
Above the upper BKT temperature $T_2$ and below the lower 
BKT temperature $T_1$, the data of different sizes begin to separate.
In the case of the BKT transition, the correlation length 
diverges as
\begin{equation}
   \xi \propto \exp (c/\sqrt{|t|}),
\label{eq:xi}
\end{equation}
where $t = T-T_{1,2}$. 

\subsection{Machine-Learning Study}
The second approach is the machine-learning study, 
developed by Shiina {\it et al.}~\cite{Shiina}, 
to classify the ordered, the BKT, and the disordered phases 
for the clock models. This work extends and generalizes 
the research conducted by Carrasquilla and Melko~\cite{Carrasquilla}, 
who focused on the Ising model. 
The configuration of a long-range spatial
correlation with the distance of $L/2$ is treated 
rather than the spin configuration itself.
In this way, a similar treatment has been given to various spin
models, including multi-component systems and systems with
a vector order parameter. The study examines not only 
the second-order and first-order transitions 
but also the BKT transition. 
For the training data, we collected typical configurations from each phase, 
the ordered, the BKT, or the disordered, 
and the test data were evaluated across all temperatures. 
We implemented a fully connected neural network using a standard library 
of TensorFlow of the 100-hidden unit model. 
A cross-entropy cost function, supplemented 
with an L2 regularization term, was employed.
The neural networks were trained using the Adam 
optimization method~\cite{Adam}.

\section{Results}

\subsection{Results of Monte Carlo Study}

\begin{figure}
\begin{center}
\includegraphics[width=6.2cm]{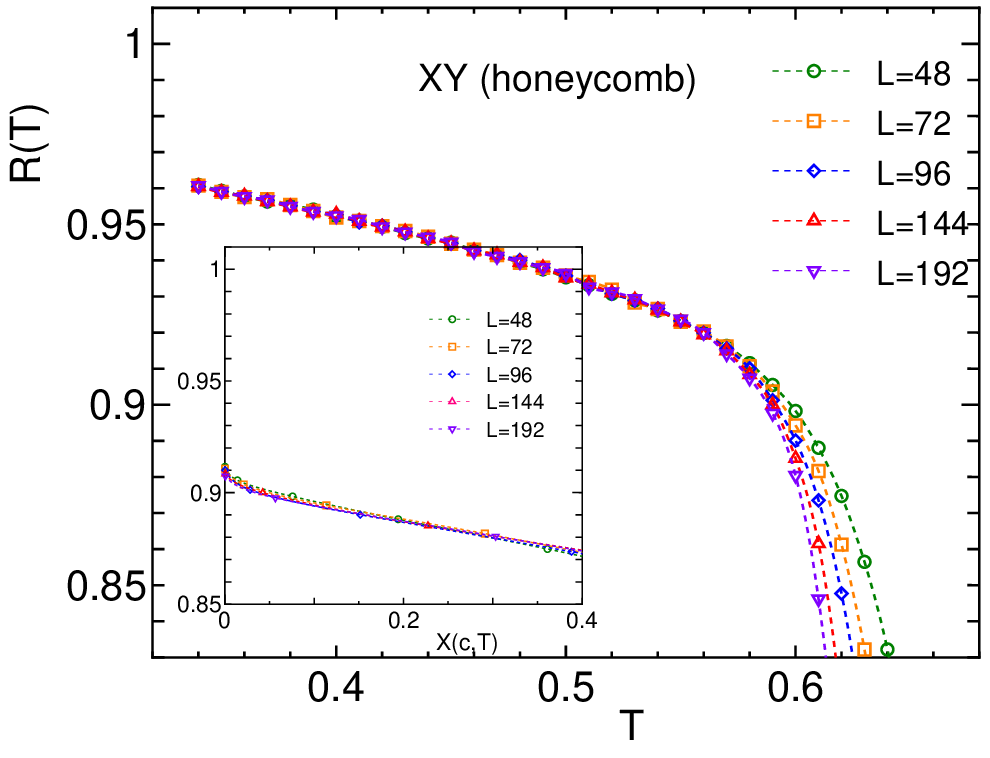}
\hspace*{4mm}
\includegraphics[width=6.2cm]{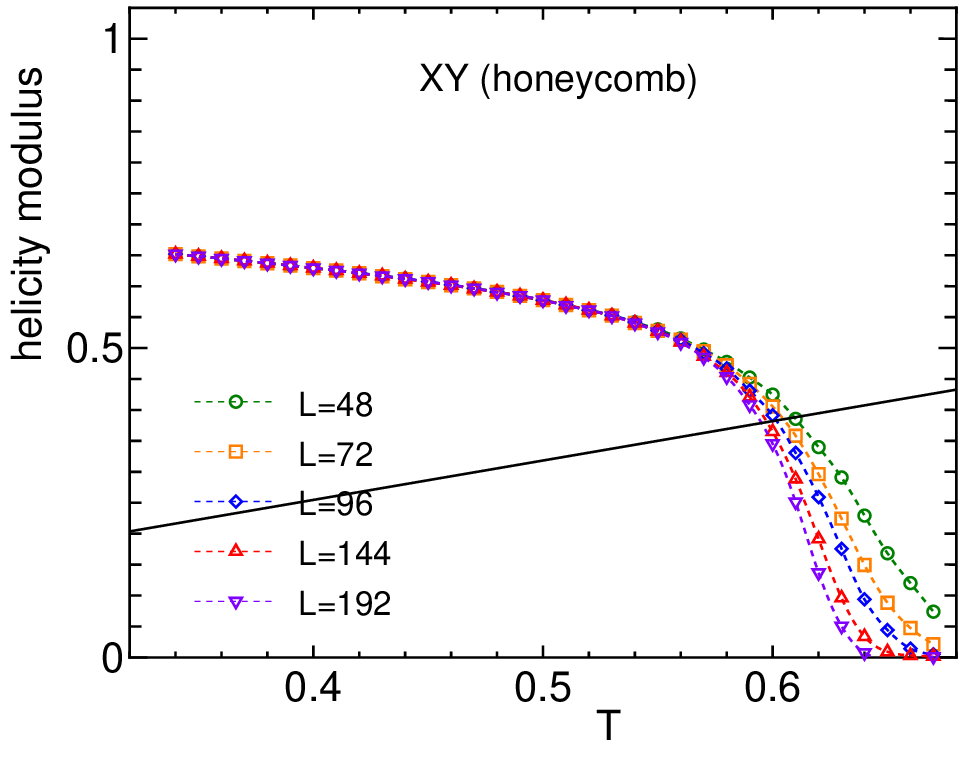}
\vspace*{2mm}
\includegraphics[width=6.2cm]{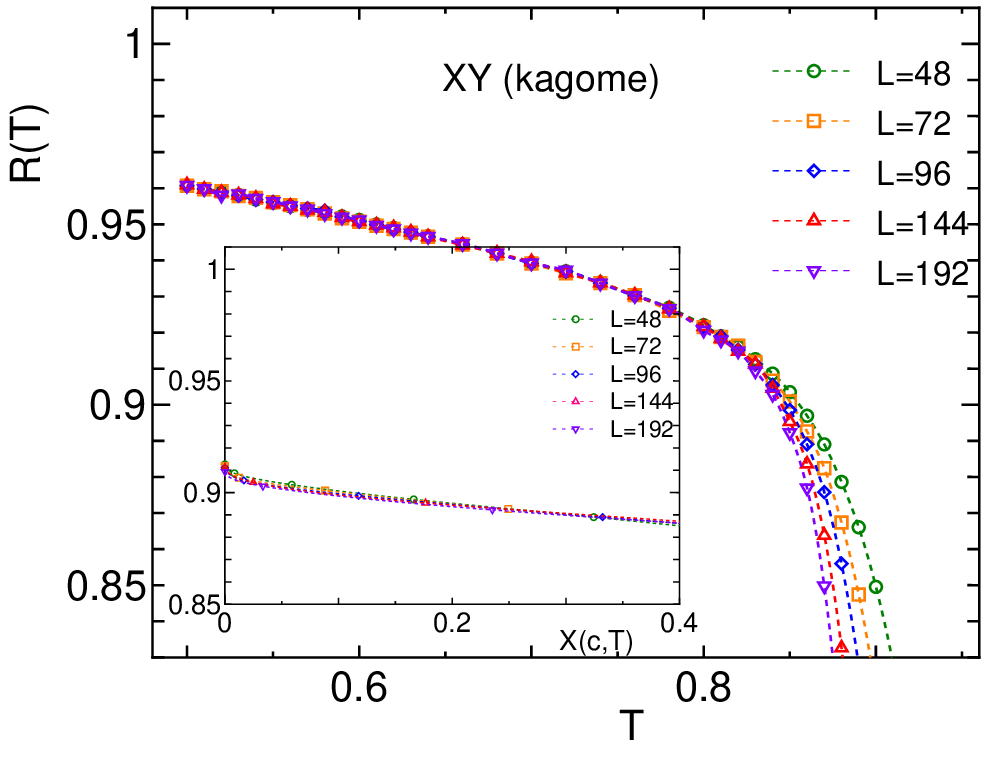}
\hspace*{4mm}
\includegraphics[width=6.2cm]{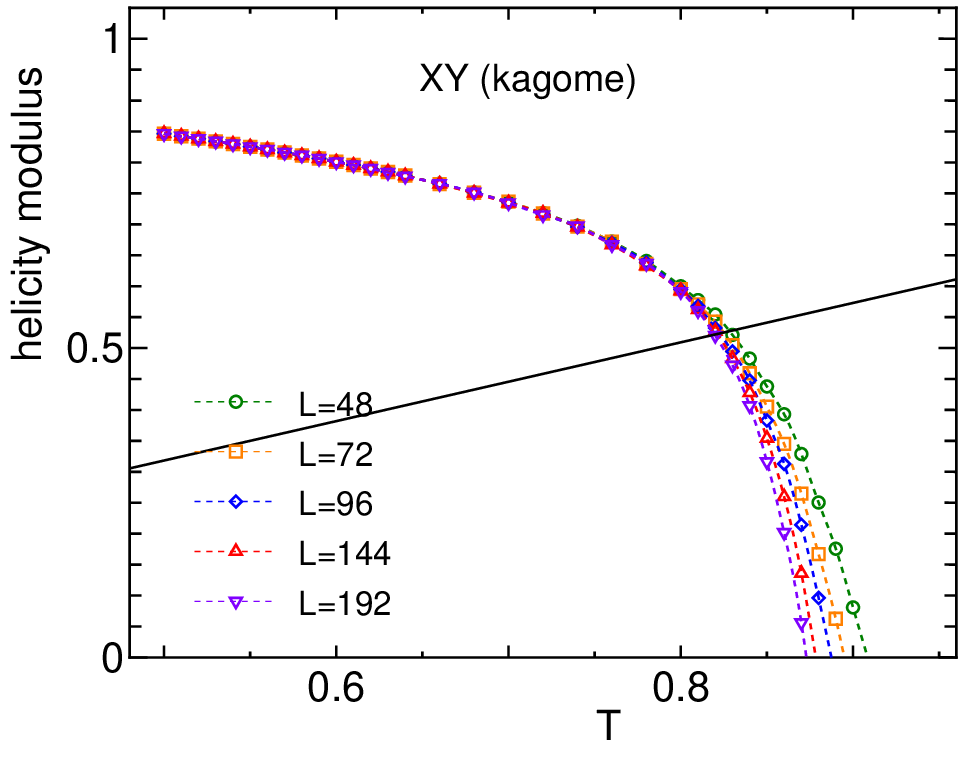}
\vspace*{2mm}
\includegraphics[width=6.2cm]{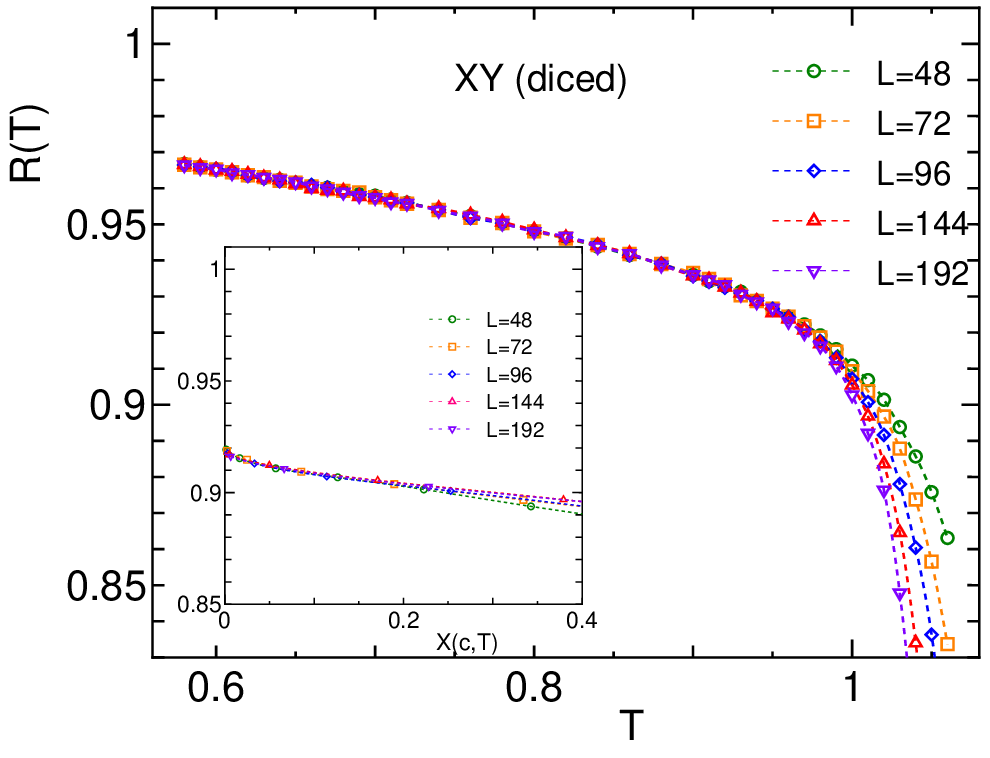}
\hspace*{4mm}
\includegraphics[width=6.2cm]{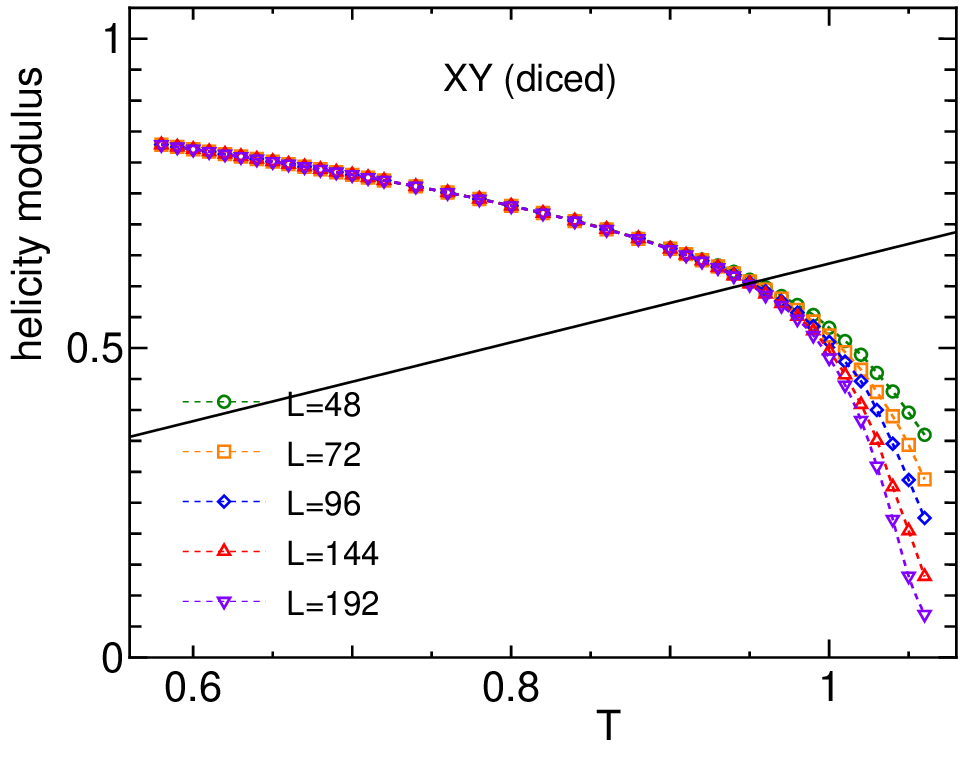}
\caption{
(left panel) The plot of the correlation ratio $R(T)$ 
of the XY model for the honeycomb, kagome, and 
diced lattice. The system sizes are 48, 72, 96, 144, and 192. 
In the inset, the FSS plots are given, where
$X(c,T)=L/\exp(c_{\rm BKT}/(\sqrt{|T-T_{\rm BKT}|}))$.
(right panel) Helicity modulus for each lattice. 
We give the straight line $(2/\pi)T$.
}
\label{fig:MC_XY}
\end{center}
\end{figure}

\subsubsection{XY model}

The Monte Carlo results of the XY model on the honeycomb,
 kagome, and diced lattices are presented in Fig.~\ref{fig:MC_XY}. 
The correlation ratio, $R(T)$, is plotted 
in the left panels of Fig.~\ref{fig:MC_XY}.  
The system sizes are $L=48$, 72, 96, 144, and 192. 

The collapse of the curves representing different system sizes 
in the plot of the correlation ratio, $R(T)$, is indicative of the BKT phase.  
The observed phenomenon is a direct consequence of the power law behavior 
exhibited by the correlation function in the BKT phase. 
We conduct a FSS analysis. 
In the inset of Fig.~\ref{fig:MC_XY}, we present the FSS plots based 
on the exponential divergence of the correlation length, as expressed 
by Eq.~(\ref{eq:xi}), 
where $X(c,T)=L/\exp(c/(|T-T_{\rm BKT}|))$, 
The rough estimate of the BKT temperature for the honeycomb lattice is 
$T_{\rm BKT}$ = 0.573.
A rough estimates of the BKT temperature for the kagome and diced lattices 
are 0.825 and 0.965, respectively.
These estimates of the BKT temperatures on the honeycomb, kagome 
and diced lattices are presented in Table~\ref{table_temperature}.

\begin{table}
\caption{
The list of BKT temperatures of the XY model and the six-state clock model 
together with the exact solutions of the Ising model. 
The second column lists the coordination numbers for each lattice. 
The exact second-order transition temperatures are provided 
in the third column, along with their values divided by the coordination number 
in the parentheses. 
The numerical estimates for the BKT transition of the XY model. 
as well as the two BKT transitions of the six-state clock model, are presented 
in the fourth, fifth, and sixth columns. Additionally, the normalized values, 
divided by the exact $T_c$ values of the Ising model, are included 
in the parentheses.
}
\begin{center}
\begin{tabular}{llllll}
      \hline
      \hline
      & \ n.n. \quad & \ $T_c^{\rm Ising}$ & \ $T_{\rm BKT}$ & \ $T_1$  & \ $T_2$ \\
      & & \ \ (/n.n.) & \ \ (/$T_c^{\rm Ising}$) & \ \ (/$T_c^{\rm Ising}$) & \ \ (/$T_c^{\rm Ising}$)\\
      \hline
      square~\cite{Hasenbusch2005,Tomita2002b}  & \ 4    & \ 2.2692  & \ 0.8929 
             & \ 0.7014   & \ 0.9008  \\
              &        & \ \ (0.567) & \ \ (0.394)& \ \ (0.309) & \ \ (0.397)\\
      triangular~\cite{Otsuka2023} & \ 6 & \ 3.6410  & \ 1.43 & \ 1.12   
                       & \ 1.44 \\
                 &     & \ \ (0.607) & \ \ (0.393)& \ \ (0.308) & \ \ (0.395)\\
      honeycomb~\cite{present}  & \ 3 & \ 1.5186  & \ 0.573 & \ 0.455   
                       & \ 0.579 \\
              &        & \ \ (0.506) & \ \ (0.377)& \ \ (0.300) & \ \ (0.381)\\
      kagome~\cite{present}  & \ 4    & \ 2.1433  & \ 0.825 & \ 0.645   
                       & \ 0.835  \\
              &        & \ \ (0.536) & \ \ (0.385) & \ \ (0.301) & \ \ (0.390)\\
      diced~\cite{present}  & \ 4 (3,6)  & \ 2.4055  & \ 0.965 & \ 0.755   
                       & \ 0.975 \\
              &        & \ \ (0.601) & \ \ (0.401) & \ \ (0.314) & \ \ (0.405)\\
      \hline
      \hline
\end{tabular}
\end{center}
\label{table_temperature}
\end{table}

The helicity modulus, $\Upsilon$, obtained by a measure of the resistance to an infinitesimal spin twist across the system along one coordinate, 
is an efficient method to calculate the BKT phase-transition temperature~\cite{Olsson}. 
Following the derivation process of Ref.~\cite{LeeDH,SunYZ}, the expression of helicity modulus can be written as
\begin{equation}
 \Upsilon = \frac{1}{L^2} \Big[ \Big\l \sum_{\l i,j \r} \cos(\theta_i-\theta_j) 
   (\hat x \cdot \hat \epsilon_{ij})^2 \Big\r
  - \frac{1}{T} \Big\l \Big( \sum_{\l i,j \r} \sin(\theta_i-\theta_j) 
   (\hat x \cdot \hat \epsilon_{ij}) \Big)^2 \Big\r \Big] .
\end{equation}
In this context, $\hat \epsilon_{ij}$ represents a unite vector pointing 
from site $j$ to site $i$. 
The symbol $\hat x$ is used to denote a selected basis vector in one coordinate.
As postulated by the renormalization-group theory~\cite{kosterlitz}, 
a universal relation exists between the helicity modulus and 
the BKT transition temperature. 
The BKT transition is distinguished by a discontinuity in the helicity modulus, 
which undergoes a transition from a value of $2T/\pi$ (in units of the Boltzmann constant $k_{\rm B}=1$)
to zero at the critical temperature. 
This critical temperature, denoted as $T_{\rm BKT}$, can be estimated 
by identifying the point of intersection between the helicity modulus $\Upsilon(T)$ and 
a straight line with the equation $\Upsilon=2T/\pi$. 

We calculate the helicity modulus, and the results for the honeycomb, kagome 
and diced lattices are plotted in the right panel of Fig.~\ref{fig:MC_XY}.
In the plot of the helicity modulus, we give the straight line $(2/\pi) T$. 
The point of intersection gives a universal jump. 
The observed intersection yields a temperature that corresponds to the BKT temperature determined by the FSS of the correlation ratio. A systematic analysis of finite size effects has been carried out~\cite{Weber,Hsieh}.

\subsubsection{six-state clock model}

\begin{figure}
\begin{center}
\includegraphics[width=6.2cm]{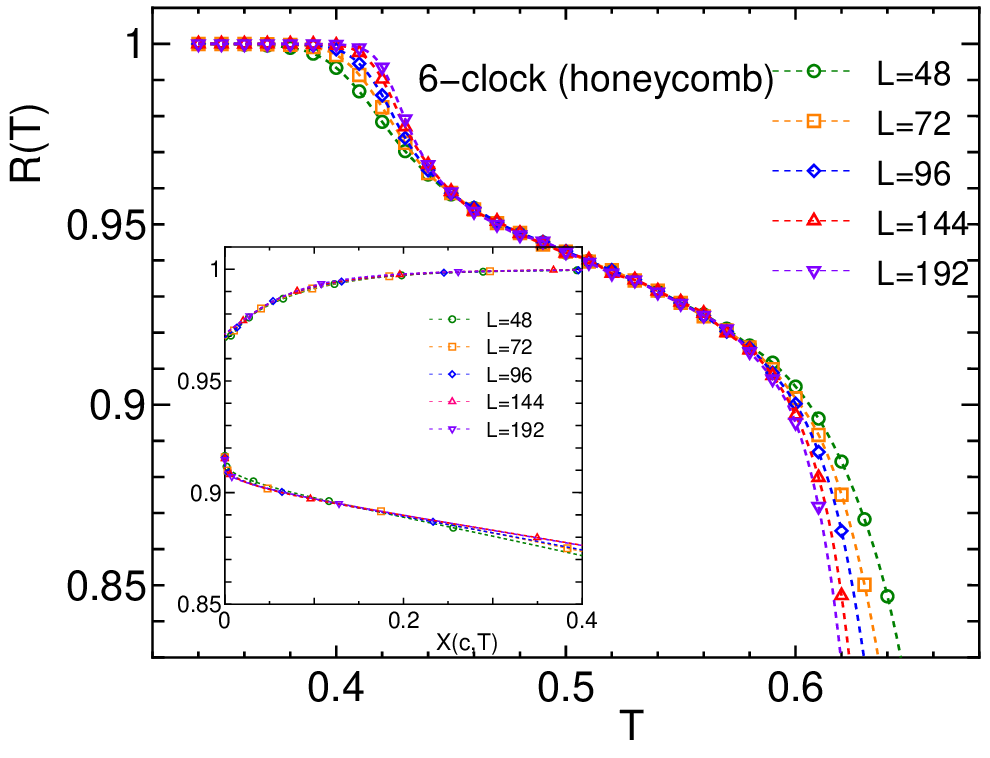}
\hspace*{4mm}
\includegraphics[width=6.2cm]{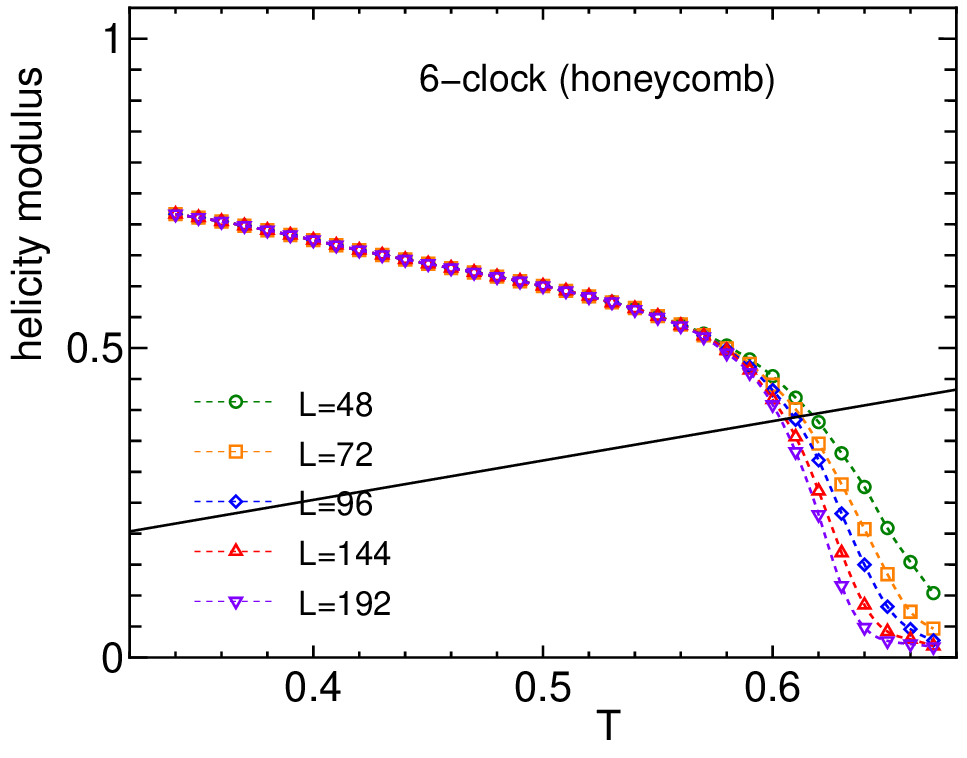}
\vspace*{2mm}
\includegraphics[width=6.2cm]{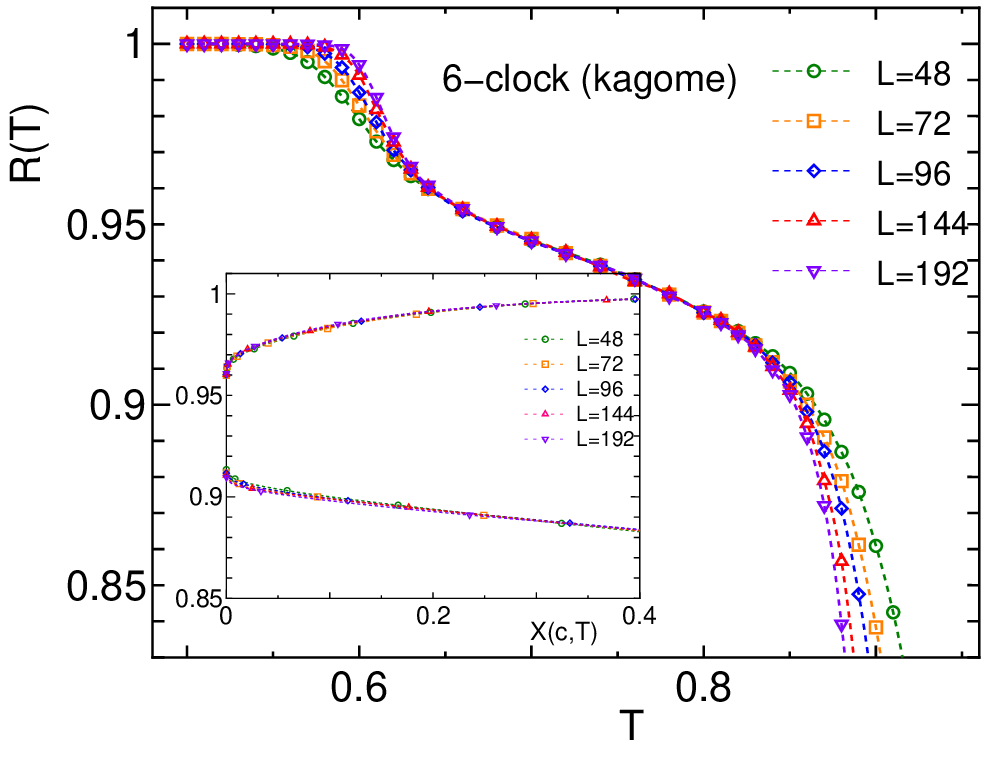}
\hspace*{4mm}
\includegraphics[width=6.2cm]{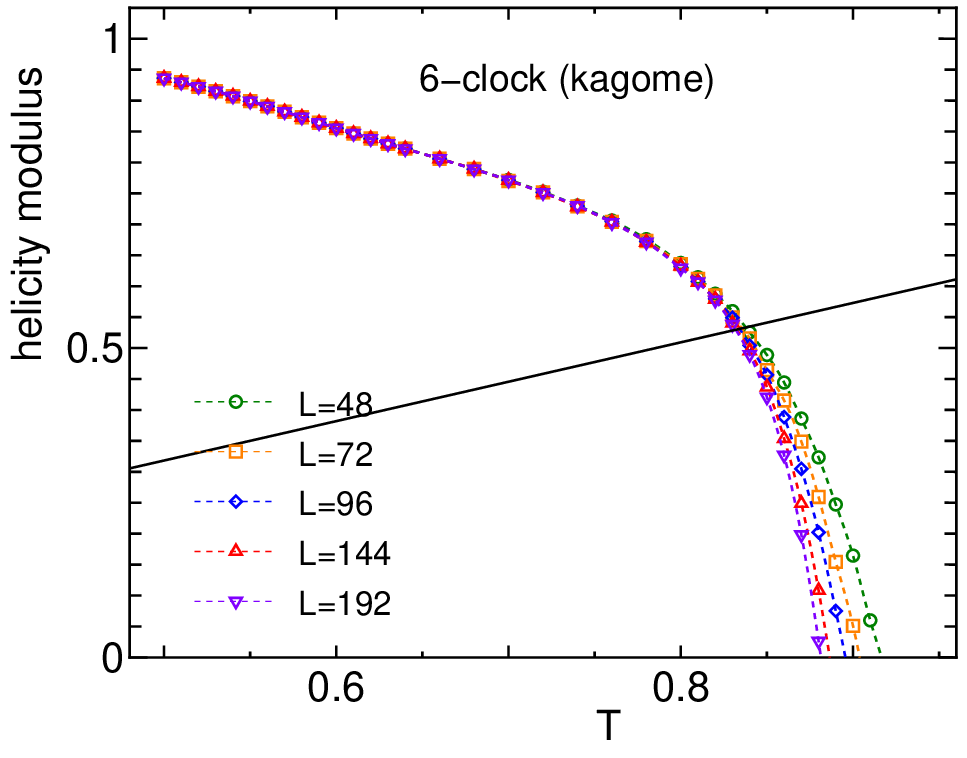}
\vspace*{2mm}
\includegraphics[width=6.2cm]{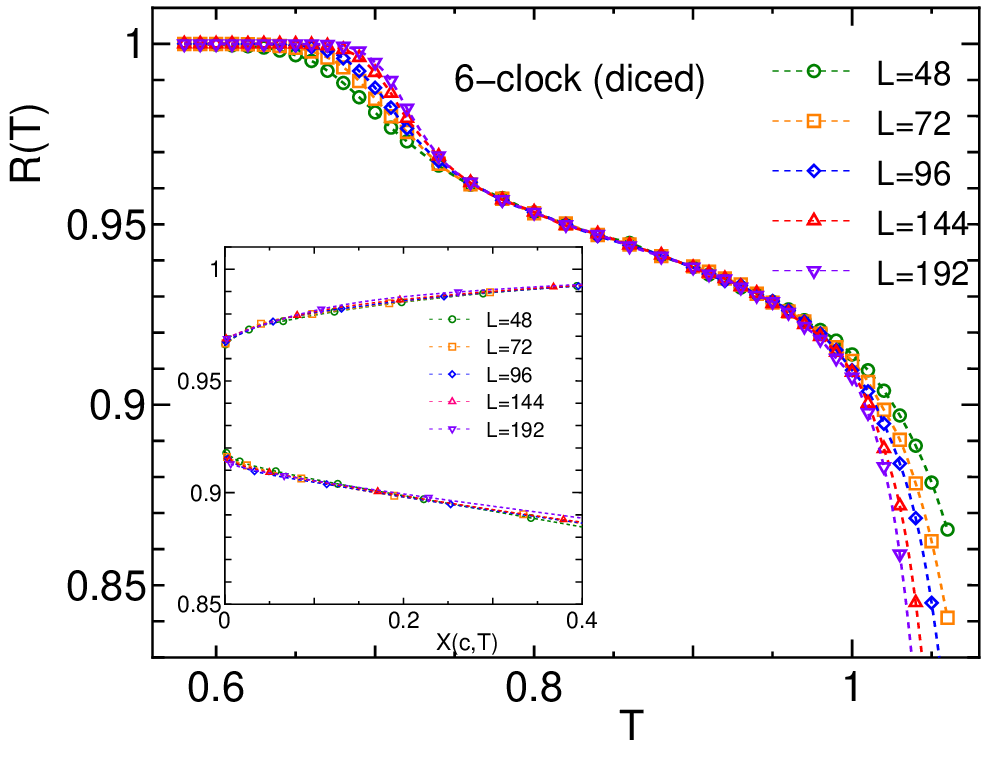}
\hspace*{4mm}
\includegraphics[width=6.2cm]{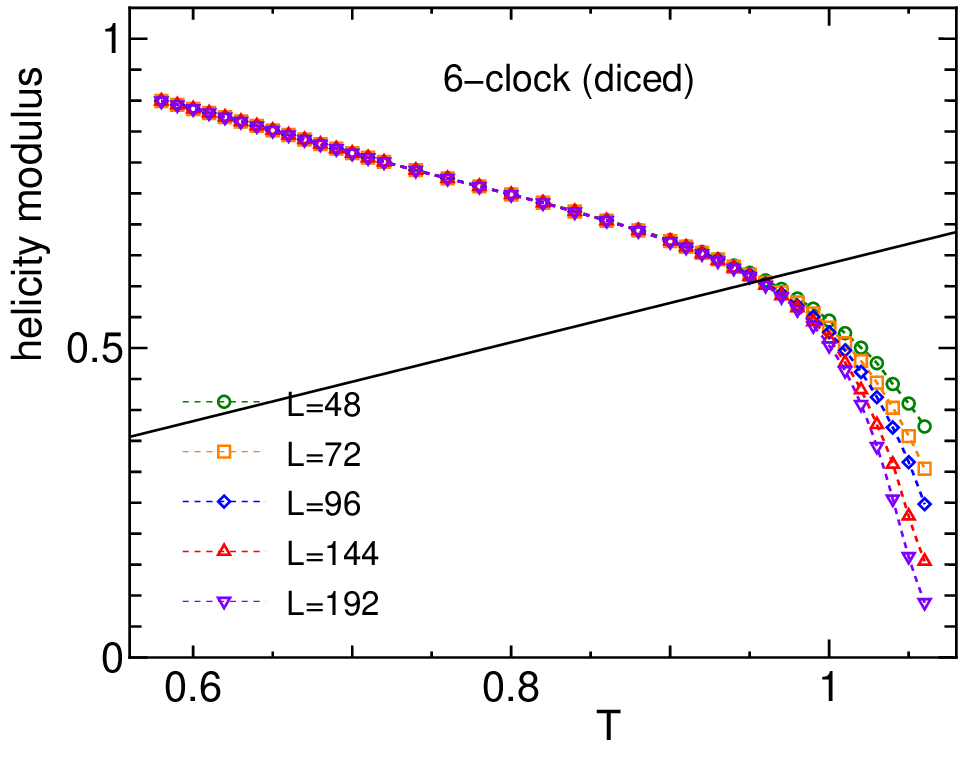}
\caption{
(left panel) The plot of the correlation ratio $R(T)$ 
of the six-state clock model for the honeycomb, kagome, and 
diced lattice. The system sizes are 48, 72, 96, 144, and 192. 
In the inset, the FSS plots are given, where
$X(c,T)=L/\exp(c_{1,2}/(\sqrt{|T-T_{1,2}|}))$.
(right panel) Helicity modulus for each lattice. 
We give the straight line $(2/\pi)T$.
}
\label{fig:MC_clock}
\end{center}
\end{figure}

The Monte Carlo results for the correlation ratio, %
$R(T)$, of the six-state clock model on the honeycomb, kagome, and 
diced lattices are presented 
in the left panel of Fig.~\ref{fig:MC_clock}. 
The system sizes are 48, 72, 96, 144, and 192. 
It can be observed that the curves for different sizes collapse
at intermediate temperature regimes 
and the spray out at lower and higher temperatures, while 
the lower and higher temperatures, a spray-out effect is evident.
The divergence of the curves at lower and higher temperatures 
indicates the occurrence of BKT transitions. 
In the inset, we present the FSS plots based on the exponential 
divergence of the correlation length, 
where $X(c,T)=L/\exp(c_{1,2}/(|T-T_{1,2}|))$. 
The BKT temperatures are estimated to be approximately 
$T_2$ = 0.579 and 
$T_1$ = 0.455 for the honeycomb lattice. 
For the kagome and diced lattices, the rough estimates are 
as follows: 
$T_2$ = 0.835, $T_1$ = 0.645
and 
$T_2$ = 0.975, $T_1$ = 0.755, 
respectively. 

The helicity modulus for the clock model was calculated and 
the results are plotted in the right panel of Fig.~\ref{fig:MC_clock}.
In the plot of helicity modulus, 
we provide the straight line $(2/\pi) T$. The crossing point gives 
a universal jump. The $q$-state clock model, which is a discrete version 
of the XY model, experiences two BKT transitions because of 
the discreteness.  We observe that there is no anomaly in helicity 
modulus for the lower transition, $T_1$. 

The behavior of the BKT transition of the high-temperature BKT 
of the six-state clock model is analogous to that observed 
in the XY model. 
The BKT temperature of the XY model for the honeycomb lattice is 
approximately $T_{\rm BKT}$ = 0.573, 
which is slightly lower than $T_2=0.579$ 
of the six-state clock model. This is similar to the square lattice case 
as well as the triangular lattice case, as shown by the previous 
studies~\cite{Tomita2002b,Otsuka2023}.
This behavior is also observed for the kagome and diced lattices.

The rough estimates of $T_1$ and $T_2$ of the six-state clock model 
on the honeycomb, kagome and diced lattices are also tabulated 
in Table~\ref{table_temperature}.

\subsection{Results of Machine-Learning Study}

\begin{figure}[t]
\begin{center}
\includegraphics[width=4.8cm]{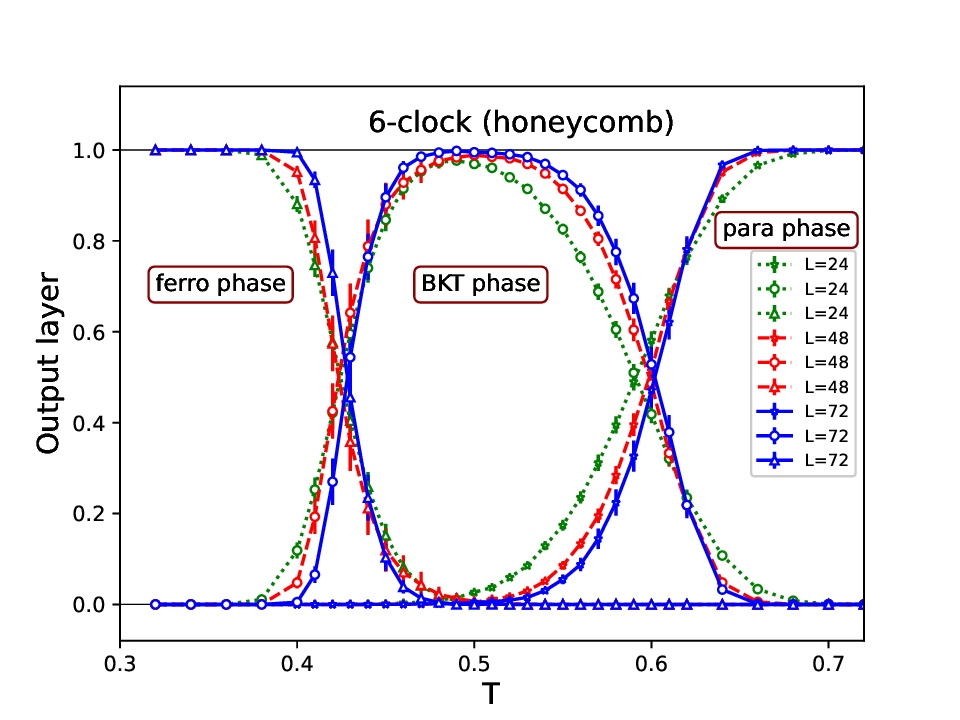}
\hspace*{2mm}
\includegraphics[width=4.8cm]{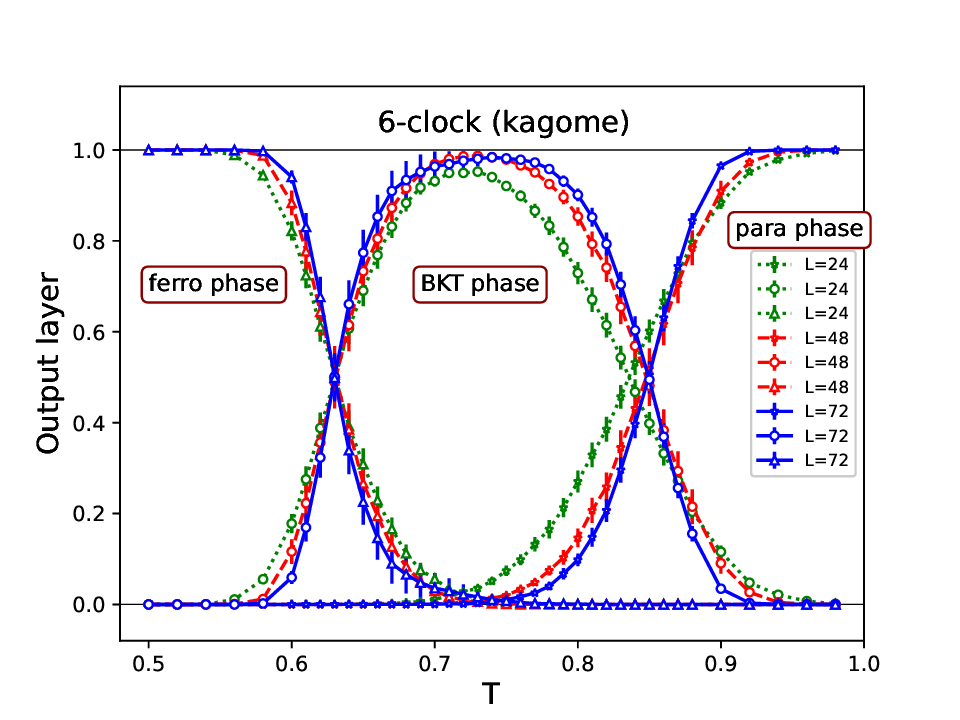}
\hspace*{2mm}
\includegraphics[width=4.8cm]{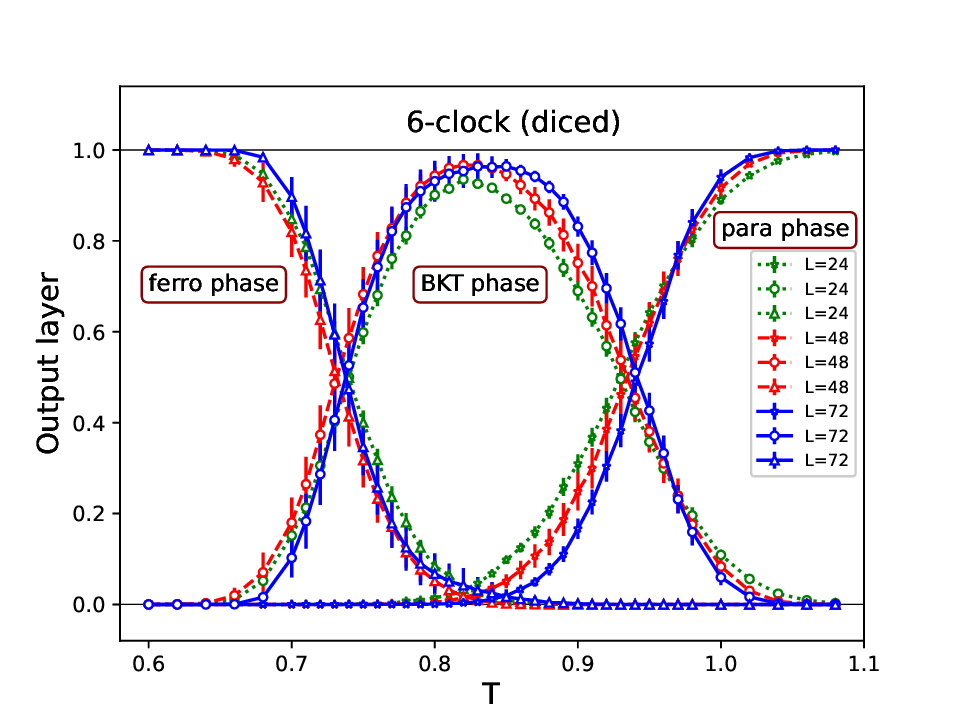}
\caption{
The machine-learning study of the six-state clock models 
on the honeycomb, kagome, and diced lattices. 
The output layer averaged over a test set as a function of $T$ 
are plotted. 
The system sizes are $L$ = 24, 48, and 72. 
For the honeycomb lattice, the samples of $T$ in the ranges 
$0.32 \le T \le 0.41$, 
$0.46 \le T \le 0.55$ and 
$0.62 \le T \le 0.71$ are used for training data. 
For the kagome lattice, the samples of $T$ in the ranges 
$0.51 \le T \le 0.60$, 
$0.69 \le T \le 0.78$ and 
$0.88 \le T \le 0.97$ are used for training data. 
For the diced lattice, the samples of $T$ in the ranges 
$0.61 \le T \le 0.70$, 
$0.79 \le T \le 0.88$ and 
$0.98 \le T \le 1.07$ are used for training data. 
}
\label{fig:output_clock6}
\end{center}
\end{figure}

The following subsection presents the findings of 
the machine-learning investigation. 
The output layers averaged over a test set as a function of $T$ 
for the six-state model on the honeycomb, kagome, and diced lattices 
are illustrated in Fig.~\ref{fig:output_clock6}. 
The system sizes are 24, 48, and 72. 
For the honeycomb lattice, the samples of $T$ in the ranges of 
$0.32 \le T \le 0.41$, $0.46 \le T \le 0.55$ and 
$0.62 \le T \le 0.71$ are used for training data, 
for the low-temperature, middle-temperature, 
and high-temperature training data, respectively.
The predicted probabilities of the phases 
are illustrated as a function of temperature. 
It is evident that three distinct phases can be discerned: 
the ordered phase, the BKT phase, and the disordered phase. 
The size-dependent $T_{1,2}(L)$ is estimated from the point 
at which the probabilities of predicting two phases are equal to 50\%. 
The approximate values of $T_1$ and $T_2$ are 0.43 
and 0.60, respectively. 
The estimates are in accordance with those derived 
from the Monte Carlo study. 
However, the BKT phase is slightly broader, which is consistent 
with the findings reported for the square lattice~\cite{Shiina}.  
This is a consequence of the finite size effect.
The size used by the machine learning method is smaller than 
that of the Monte Carlo method. Additionally, the logarithmic correction might be less significant for the Monte Carlo method, which deals with correlation ratios.

For the kagome lattice, the samples of $T$ in the ranges 
$0.51 \le T \le 0.60$, $0.69 \le T \le 0.78$ and $0.88 \le T \le 0.97$ 
are employed as training data. 
In accordance with the methodology previously employed 
for the honeycomb lattice, 
the rough estimates of $T_1$ and $T_2$ are 0.63 
and 0.85, respectively. 
These estimates are once again found to be compatible 
with those obtained from the Monte Carlo study. 
In the case of the diced lattice, the training data set is comprised 
of samples of $T$ in the ranges 
$0.61 \le T \le 0.70$, $0.79 \le T \le 0.88$ and $0.98 \le T \le 1.07$. 
The approximate values of $T_1$ and $T_2$ are 0.74 
and 0.94, respectively. 

\section{Comparison of BKT temperatures}

Exact solutions of the Ising models on various 2D lattices 
have been obtained.  The square-lattice Ising model was exactly 
solved by Onsager~\cite{Onsager}. 
The exact solutions for the transition temperatures have also 
been obtained for the triangular and honeycomb lattices \cite{Husimi}, 
and for the kagome and diced lattices \cite{Syozi, Kano}. 
The critical temperatures of the second-order transition 
for these lattices are
\begin{eqnarray*}
  T_c^{\rm square} &=& \frac{2}{\ln(1+\sqrt{2})} = 2.26919 \cdots \\
  T_c^{\rm triangular} &=& \frac{4}{\ln 3} = 3.640964 \cdots \\
  T_c^{\rm honeycomb} &=& \frac{2}{\cosh^{-1}(2)}=1.5186 \cdots \\
  T_c^{\rm kagome} &=& \frac{4}{\ln(3+2\sqrt{3})}=2.143319 \cdots \\
  T_c^{\rm diced} &=& \frac{2}{\cosh^{-1}\frac{1+\sqrt{3}}{2}}
  =2.405457 \cdots.
\end{eqnarray*}

Here, we should mention the concept of duality in lattices. 
The triangular lattice and the honeycomb lattice are dual to each other, 
as are the kagome lattice and the diced lattice. 
Additionally, the square lattice is self-dual. 
Kramers and Wannier~\cite{Kramers,Syozi72} demonstrated 
that there is a duality relation 
between the Ising models of these dual lattices,  
and they established the relationship between their transition temperatures 
as follows:
\begin{equation}
  \sinh(2J/T_{\rm c})\sinh(2J/T_{\rm c}^*) = 1.
\end{equation}

The evaluated BKT transition temperatures for the XY and six-state clock models 
are summarized in Table~\ref{table_temperature}. 
For each lattice, a comparison is made with the second-order 
phase transition temperatures of the Ising model. 
The second column of Table~\ref{table_temperature} presents 
the coordination number for each lattice. In the case of the diced lattice, 
the coordination number can be either 3 or 6; however, the average value 
of 4 is indicated in parentheses. The third column provides 
the second-order phase transition temperature of the Ising model, 
divided by the coordination number (also in parentheses). 
The similarity in these values suggests that the transition temperature 
is approximately proportional to the coordination number. 

The fourth column gives the BKT transition temperature of the XY model, 
divided by the transition temperature of the Ising model, 
with the results shown in parentheses. This value is consistently 
around 0.37 to 0.40, indicating the independence from the lattice structure. 
The two BKT transition temperatures of the six-state clock model, 
denoted as $T_1$ and $T_2$, are presented in the fifth and sixth columns, 
respectively. Values for these temperatures, normalized by $T_c$ 
of the Ising model are also included in parentheses. 
Again, this normalized value appears to be approximately constant, 
regardless of the lattice type.

\section{Summary and discussion}

Monte Carlo and machine learning methods were systematically employed to investigate XY models and six-state clock models on honeycomb, kagome, and diced lattices. The study examined the BKT transition temperature, in conjunction with the results of previous research focused on triangular lattices.

The Monte Carlo method was used to accurately estimate the BKT transition temperature by analyzing the ratio of correlations at different distances, which helped reduce the numerical difficulties associated with logarithmic divergence. 
Through FSS analysis, we identified the BKT temperature of the XY model and 
the two BKT temperatures of the six-state clock model. 
Additionally, machine learning techniques~\cite{Shiina} extended the approach~\cite{Shiina} by Carrasquilla and Melko~\cite{Carrasquilla}, further validating its effectiveness in classifying low-temperature ordered, intermediate BKT, and high-temperature disordered phases by examining configurations of long-range correlations. The estimated BKT temperatures allign well with those obtained from the Monte Carlo study.

In discussing the BKT transition across various 2D lattices, we have compared our results to those of the Ising model, which has known exact solutions. The BKT temperature of the XY model, along with the two BKT transition temperatures $T_1$ and $T_2$ of the six-state clock model, can be normalized by the transition temperature of the Ising model. Although the normalized values are not universal, they are nearly equal, which is a logical conclusion.

Recently, two research groups~\cite{Andrade,Jiang} have reported findings on the XY model of the honeycomb lattice, and their estimates are 
0.576(1)\cite{Andrade} and 0.560(9) \cite{Jiang}.
The present calculation, 0.573, is consistent with both results. Some literature has pointed out a difference from Nienhuis's exact solution~\cite{Nienhuis,Wang_Zhang}; however, since Nienhuis's model is also referred to as the $n$-vector loop model and involves different interactions. Therefore, it is expected that the numerical results from our study and the two previous papers would differ from Nienhuis's exact result.

Research using machine learning techniques has investigated the six-state clock universality of antiferromagnetic systems with next-nearest neighbor interactions, specifically focusing on a triangular lattice~\cite{Otsuka2023}. 
In this study, the data from the ferromagnetic six-state clock model on the triangular lattice served as training data. 
The antiferromagnetic Ising model on the kagome lattice with next-nearest-neighbor ferromagnetic interactions has been analyzed for quite some time and represents a unique system that demonstrates six-state clock universality~\cite{Takagi,GiaWei}. While earlier calculations were not sufficiently quantitative, recent reports have offered new insights~\cite{Su23}. It will be intriguing to apply machine learning techniques to this model, especially now that we have data on the six-state ferromagnetic clock model on the kagome lattice. 

We comment on the logarithmic corrections in the FSS. There are multiplicative logarithmic corrections in the amplitude of the FSS of the susceptibility~\cite{kosterlitz2,Janke}. Accurately reproducing the theoretically predicted value of the exponent that specifies this multiplicative logarithmic correction has proven to be a challenging numerical problem~\cite{Komura}. 
It is interesting to explore whether the multiplicative logarithmic correction exponent is more straightforward to reproduce for other lattices.

The problem of the extraordinary-log surface phase transition 
for the three-dimensional (3D) XY model is a new topic 
in the XY universality~\cite{HuM}. 
The surface of 3D systems is two-dimensional, and the surface effects 
are classified into ordinary, special, and extraordinary transitions 
depending on the strength of the surface interaction. 
In the extraordinary transition of the XY system, it has been pointed out that the correlation function does not behave in a simple power-law manner but in a logarithmic manner. How the extraordinary-log surface phase transitions behave in surface forms of different lattices instead of cubic lattices will be the subject of future research.

\section*{Data availability statement}

All data that support the findings of this study are included within the article.

\section*{Acknowledgment}

The authors would like to thank Kenta Shiina, Hiroyuki Mori, and Hwee Kuan Lee 
for valuable discussions.  
This work was supported by JSPS KAKENHI Grant Number JP22K03472. 

\newpage
\newcommand{\mybibitem}[6]{\bibitem{#1} #2 #6 #3 {\it #4} #5}

\end{document}